\input harvmac
\input amssym
%\draftmode

\def\p{\partial}

\def\rt{\rightarrow}
\def\Oc{{\cal O}}

\def\av{\vec{a}}

\def\zb{\overline{z}}

\def\taub{\overline{\tau}}

\def\Ab{\overline{A}}
\def\Fb{\overline{F}}

\def\zb{\overline{z}}
\def\Lc{{\cal L}}
\def\Lcb{\overline{\cal L}}

\def\Wc{{\cal W}}
\def\Wcb{{\overline{\cal W}}}

\def\lamb{\overline{\lambda}}

\def\ab{\overline{a}}
\def\omb{\overline{\omega}}
\def\mub{\overline{\mu}}

\def\cb{\overline{c}}

\def\Hb{\overline{H}}
\def\omb{\overline{\omega}}
\def\wb{\overline{w}}
\def\ellb{\overline{\ell}}
\def\alphab{\overline{\alpha}}

%\WittenHC
\lref\WittenHC{
  E.~Witten,
  ``(2+1)-Dimensional Gravity as an Exactly Soluble System,''
  Nucl.\ Phys.\  B {\bf 311}, 46 (1988).
  %%CITATION = NUPHA,B311,46;%%
}

%\AchucarroVZ
\lref\AchucarroVZ{
  A.~Achucarro and P.~K.~Townsend,
  ``A Chern-Simons Action for Three-Dimensional anti-De Sitter Supergravity
  Theories,''
  Phys.\ Lett.\  B {\bf 180}, 89 (1986).
  %%CITATION = PHLTA,B180,89;%%
}

%\WittenKT
\lref\WittenKT{
  E.~Witten,
  ``Three-Dimensional Gravity Revisited,''
  arXiv:0706.3359 [hep-th].
  %%CITATION = ARXIV:0706.3359;%%
}

%\BalasubramanianRE
\lref\BalasubramanianRE{
  V.~Balasubramanian and P.~Kraus,
  ``A stress tensor for anti-de Sitter gravity,''
  Commun.\ Math.\ Phys.\  {\bf 208}, 413 (1999)
  [arXiv:hep-th/9902121].
  %%CITATION = CMPHA,208,413;%%
}

%\KrausNB
\lref\KrausNB{
  P.~Kraus and F.~Larsen,
  ``Partition functions and elliptic genera from supergravity,''
  JHEP {\bf 0701}, 002 (2007)
  [arXiv:hep-th/0607138].
  %%CITATION = JHEPA,0701,002;%%
}

%%%%%%%%%%%%%%%%%%%%

%\PolchinskiRQ
\lref\PolchinskiRQ{
  J.~Polchinski,
  ``String theory. Vol. 1: An introduction to the bosonic string,''
%\href{http://www.slac.stanford.edu/spires/find/hep/www?irn=4634799}{SPIRES entry}
{\it  Cambridge, UK: Univ. Pr. (1998) 402 p}
}

%\BrownNW
\lref\BrownNW{
  J.~D.~Brown and M.~Henneaux,
  ``Central Charges in the Canonical Realization of Asymptotic Symmetries: An
  Example from Three-Dimensional Gravity,''
  Commun.\ Math.\ Phys.\  {\bf 104}, 207 (1986).
  %%CITATION = CMPHA,104,207;%%
}

%\DijkgraafFQ
\lref\DijkgraafFQ{
  R.~Dijkgraaf, J.~M.~Maldacena, G.~W.~Moore and E.~P.~Verlinde,
  ``A black hole farey tail,''
  arXiv:hep-th/0005003.
  %%CITATION = HEP-TH/0005003;%%
}

%%%%%%%%%%%%%%%%%%%%%%%%%%%%%%%%%%

%\HaggiManiRU
\lref\HaggiManiRU{
  P.~Haggi-Mani and B.~Sundborg,
  ``Free large N supersymmetric Yang-Mills theory as a string theory,''
  JHEP {\bf 0004}, 031 (2000)
  [arXiv:hep-th/0002189].
  %%CITATION = JHEPA,0004,031;%%
}

%\KonsteinBI
\lref\KonsteinBI{
  S.~E.~Konstein, M.~A.~Vasiliev and V.~N.~Zaikin,
  ``Conformal higher spin currents in any dimension and AdS/CFT
  correspondence,''
  JHEP {\bf 0012}, 018 (2000)
  [arXiv:hep-th/0010239].
  %%CITATION = JHEPA,0012,018;%%
}

%\SundborgWP
\lref\SundborgWP{
  B.~Sundborg,
  ``Stringy gravity, interacting tensionless strings and massless higher
  spins,''
  Nucl.\ Phys.\ Proc.\ Suppl.\  {\bf 102}, 113 (2001)
  [arXiv:hep-th/0103247].
  %%CITATION = NUPHZ,102,113;%%
}

%\MikhailovBP
\lref\MikhailovBP{
  A.~Mikhailov,
  ``Notes on higher spin symmetries,''
  arXiv:hep-th/0201019.
  %%CITATION = HEP-TH/0201019;%%
}

%\SezginRT
\lref\SezginRT{
  E.~Sezgin and P.~Sundell,
  ``Massless higher spins and holography,''
  Nucl.\ Phys.\  B {\bf 644}, 303 (2002)
  [Erratum-ibid.\  B {\bf 660}, 403 (2003)]
  [arXiv:hep-th/0205131].
  %%CITATION = NUPHA,B644,303;%%
}

%\KlebanovJA
\lref\KlebanovJA{
  I.~R.~Klebanov and A.~M.~Polyakov,
  ``AdS dual of the critical O(N) vector model,''
  Phys.\ Lett.\  B {\bf 550}, 213 (2002)
  [arXiv:hep-th/0210114].
  %%CITATION = PHLTA,B550,213;%%
}

%\GiombiWH
\lref\GiombiWH{
  S.~Giombi and X.~Yin,
  ``Higher Spin Gauge Theory and Holography: The Three-Point Functions,''
  JHEP {\bf 1009}, 115 (2010)
  [arXiv:0912.3462 [hep-th]].
  %%CITATION = JHEPA,1009,115;%%
}

%\GiombiVG
\lref\GiombiVG{
  S.~Giombi and X.~Yin,
  ``Higher Spins in AdS and Twistorial Holography,''$\quad \quad$
  arXiv:1004.3736 [hep-th].
  %%CITATION = ARXIV:1004.3736;%%
}

%\HenneauxXG
\lref\HenneauxXG{
  M.~Henneaux and S.~J.~Rey,
  ``Nonlinear W(infinity) Algebra as Asymptotic Symmetry of Three-Dimensional
  Higher Spin Anti-de Sitter Gravity,''
  JHEP {\bf 1012}, 007 (2010)
  [arXiv:1008.4579 [hep-th]].
  %%CITATION = JHEPA,1012,007;%%
}

%\CampoleoniZQ
\lref\CampoleoniZQ{
  A.~Campoleoni, S.~Fredenhagen, S.~Pfenninger and S.~Theisen,
  ``Asymptotic symmetries of three-dimensional gravity coupled to higher-spin
  fields,''
  JHEP {\bf 1011}, 007 (2010)
  [arXiv:1008.4744 [hep-th]].
  %%CITATION = JHEPA,1011,007;%%
}

%\GaberdielAR
\lref\GaberdielAR{
  M.~R.~Gaberdiel, R.~Gopakumar and A.~Saha,
  ``Quantum W-symmetry in AdS$_3$,''
  JHEP {\bf 1102}, 004 (2011)
  [arXiv:1009.6087 [hep-th]].
  %%CITATION = JHEPA,1102,004;%%
}

%\GaberdielPZ
\lref\GaberdielPZ{
  M.~R.~Gaberdiel and R.~Gopakumar,
  ``An AdS$_3$ Dual for Minimal Model CFTs,''
  arXiv:1011.2986 [hep-th].
  %%CITATION = ARXIV:1011.2986;%%
}

%\DouglasRC
\lref\DouglasRC{
  M.~R.~Douglas, L.~Mazzucato and S.~S.~Razamat,
  ``Holographic dual of free field theory,''
  arXiv:1011.4926 [hep-th].
  %%CITATION = ARXIV:1011.4926;%%
}

%\CastroCE
\lref\CastroCE{
  A.~Castro, A.~Lepage-Jutier and A.~Maloney,
  ``Higher Spin Theories in AdS$_3$ and a Gravitational Exclusion Principle,''
  JHEP {\bf 1101}, 142 (2011)
  [arXiv:1012.0598 [hep-th]].
  %%CITATION = JHEPA,1101,142;%%
}

%\GaberdielWB
\lref\GaberdielWB{
  M.~R.~Gaberdiel and T.~Hartman,
  ``Symmetries of Holographic Minimal Models,''
  arXiv:1101.2910 [hep-th].
  %%CITATION = ARXIV:1101.2910;%%
}

%%%%%%%%%%%%%%%%%%%%%%%%%%%%%%%%

%\BoulangerDD
\lref\BoulangerDD{
  N.~Boulanger and P.~Sundell,
  ``An action principle for Vasiliev's four-dimensional higher-spin gravity,''
  arXiv:1102.2219 [hep-th].
  %%CITATION = ARXIV:1102.2219;%%
}

%\DoroudXS
\lref\DoroudXS{
  N.~Doroud and L.~Smolin,
  ``An action for higher spin gauge theory in four dimensions,''
  arXiv:1102.3297 [hep-th].
  %%CITATION = ARXIV:1102.3297;%%
}

%%%%%%%%%%%%%%%%%%%%%%%%%%%%%%%%

%\FradkinKS
\lref\FradkinKS{
  E.~S.~Fradkin and M.~A.~Vasiliev,
  ``On the Gravitational Interaction of Massless Higher Spin Fields,''
  Phys.\ Lett.\  B {\bf 189}, 89 (1987).
  %%CITATION = PHLTA,B189,89;%%
}

%\FradkinQY
\lref\FradkinQY{
  E.~S.~Fradkin and M.~A.~Vasiliev,
  ``Cubic Interaction in Extended Theories of Massless Higher Spin Fields,''
  Nucl.\ Phys.\  B {\bf 291}, 141 (1987).
  %%CITATION = NUPHA,B291,141;%%
}

%\VasilievTK
\lref\VasilievTK{
  M.~A.~Vasiliev,
  ``Equations of motion for interacting massless fields of all spins in
  (3+1)-dimensions,''
%\href{http://www.slac.stanford.edu/spires/find/hep/www?irn=3033953}{SPIRES entry}
{\it  In *Moscow 1990, Proceedings, Symmetries and algebraic structures in physics, pt. 1* 15-33}
}

%\VasilievAV
\lref\VasilievAV{
  M.~A.~Vasiliev,
  ``More On Equations Of Motion For Interacting Massless Fields Of All Spins In
  (3+1)-Dimensions,''
  Phys.\ Lett.\  B {\bf 285}, 225 (1992).
  %%CITATION = PHLTA,B285,225;%%
}

%\BlencoweGJ
\lref\BlencoweGJ{
  M.~P.~Blencowe,
  ``A Consistent Interacting Massless Higher Spin Field Theory In D = (2+1),''
  Class.\ Quant.\ Grav.\  {\bf 6}, 443 (1989).
  %%CITATION = CQGRD,6,443;%%
}

%\BergshoeffNS
\lref\BergshoeffNS{
  E.~Bergshoeff, M.~P.~Blencowe and K.~S.~Stelle,
  ``Area Preserving Diffeomorphisms And Higher Spin Algebra,''
  Commun.\ Math.\ Phys.\  {\bf 128}, 213 (1990).
  %%CITATION = CMPHA,128,213;%%
}

%\StromingerEQ
\lref\StromingerEQ{
  A.~Strominger,
  ``Black hole entropy from near-horizon microstates,''
  JHEP {\bf 9802}, 009 (1998)
  [arXiv:hep-th/9712251].
  %%CITATION = JHEPA,9802,009;%%
}

%\ZamolodchikovWN
\lref\ZamolodchikovWN{
  A.~B.~Zamolodchikov,
  ``Infinite Additional Symmetries In Two-Dimensional Conformal Quantum Field
  Theory,''
  Theor.\ Math.\ Phys.\  {\bf 65}, 1205 (1985)
  [Teor.\ Mat.\ Fiz.\  {\bf 65}, 347 (1985)].
  %%CITATION = TMFZA,65,347;%%
}

%\KrausWN
\lref\KrausWN{
  P.~Kraus,
  ``Lectures on black holes and the AdS(3)/CFT(2) correspondence,''
  Lect.\ Notes Phys.\  {\bf 755}, 193 (2008)
  [arXiv:hep-th/0609074].
  %%CITATION = LNPHA,755,193;%%
}

%\HenneauxIB
\lref\HenneauxIB{
  M.~Henneaux, L.~Maoz and A.~Schwimmer,
  ``Asymptotic dynamics and asymptotic symmetries of three-dimensional
  extended AdS supergravity,''
  Annals Phys.\  {\bf 282}, 31 (2000)
  [arXiv:hep-th/9910013].
  %%CITATION = APNYA,282,31;%%
}

%\LapanJX
\lref\LapanJX{
  J.~M.~Lapan, A.~Simons and A.~Strominger,
  ``Nearing the Horizon of a Heterotic String,''
  arXiv:0708.0016 [hep-th].
  %%CITATION = ARXIV:0708.0016;%%
}

%\KrausVU
\lref\KrausVU{
  P.~Kraus, F.~Larsen and A.~Shah,
  ``Fundamental Strings, Holography, and Nonlinear Superconformal Algebras,''
  JHEP {\bf 0711}, 028 (2007)
  [arXiv:0708.1001 [hep-th]].
  %%CITATION = JHEPA,0711,028;%%
}

%\BanadosNR
\lref\BanadosNR{
  M.~Banados and R.~Caro,
  ``Holographic Ward identities: Examples from 2+1 gravity,''
  JHEP {\bf 0412}, 036 (2004)
  [arXiv:hep-th/0411060].
  %%CITATION = JHEPA,0412,036;%%
}

%\BanadosGG
\lref\BanadosGG{
  M.~Banados,
  ``Three-dimensional quantum geometry and black holes,''
  arXiv:hep-th/9901148.
  %%CITATION = HEP-TH/9901148;%%
}

%\FronsdalRB
\lref\FronsdalRB{
  C.~Fronsdal,
  ``Massless Fields With Integer Spin,''
  Phys.\ Rev.\  D {\bf 18}, 3624 (1978).
  %%CITATION = PHRVA,D18,3624;%%
}

%\BouwknegtNG
\lref\BouwknegtNG{
  P.~Bouwknegt, J.~G.~McCarthy and K.~Pilch,
  ``Semiinfinite cohomology of W algebras,''
  Lett.\ Math.\ Phys.\  {\bf 29}, 91 (1993)
  [arXiv:hep-th/9302086].
  %%CITATION = LMPHD,29,91;%%
}

%\DidenkoTD
\lref\DidenkoTD{
  V.~E.~Didenko and M.~A.~Vasiliev,
  ``Static BPS black hole in 4d higher-spin gauge theory,''
  Phys.\ Lett.\  B {\bf 682}, 305 (2009)
  [arXiv:0906.3898 []].
  %%CITATION = PHLTA,B682,305;%%
}

%\KochCY
\lref\KochCY{
  R.~d.~M.~Koch, A.~Jevicki, K.~Jin and J.~P.~Rodrigues,
  ``$AdS_4/CFT_3$ Construction from Collective Fields,''
  Phys.\ Rev.\  D {\bf 83}, 025006 (2011)
  [arXiv:1008.0633].
  %%CITATION = PHRVA,D83,025006;%%
}

%\AmmonNK
\lref\AmmonNK{
  M.~Ammon, M.~Gutperle, P.~Kraus and E.~Perlmutter,
  ``Spacetime Geometry in Higher Spin Gravity,''
  arXiv:1106.4788 [hep-th].
  %%CITATION = ARXIV:1106.4788;%%
}

%%%%%%%%%%%%%%%%%%%%%%%%%%%%%%%%%%%%%%%%%%%%%%%%%%%%%%%%%%%%%%%%%%%%%%%%%

\Title{\vbox{\baselineskip12pt
%\hbox{hep-th/0508218}
%\hbox{UCLA-05-TEP-XX} \hbox{MCTP-XX-XX}
}} {\vbox{\centerline {Higher Spin Black Holes}}}
%\medskip\vbox{\centerline {More title}}} }
\centerline{
  Michael Gutperle$^\dagger$\foot{gutperle@ucla.edu} and Per
Kraus$^{\dagger}$\foot{pkraus@ucla.edu}}
\bigskip
\centerline{${}^\dagger$\it{Department of Physics and Astronomy}}
\centerline{${}$\it{University of California, Los Angeles, CA 90095,USA}}

\baselineskip16pt

\vskip .3in

\centerline{\bf Abstract}

We study classical solutions of  three dimensional higher spin gravity in the Chern-Simons formulation. We  find  solutions that generalize the BTZ black hole and  carry spin-3 charge.  The black hole entropy formula yields a result for the asymptotic growth
of the partition function at finite spin-3 chemical potential.  Along the way, we develop technology for computing AdS/CFT correlation functions involving higher spin operators.

%%%
\Date{March, 2011}
%%%%%%%%%%%%%%%%%%%%%%%%%%%%%%%%%%%%%%%%%%%%%%
%%%%%%%%%%%%%%%%%%%%%%%%%%%
% Main text begins here
%%%%%%%%%%%%%%%%%%%%%%%%%%%%%%%%%%%%%%%%%%%%%%
%%%%%%%%%%%%%%%%%%%%%%%%%%%
\baselineskip16pt

%\listtoc \writetoc

\newsec{Introduction}

Theories of gravity coupled to massless higher spin fields \refs{\FronsdalRB,\FradkinKS,\FradkinQY,\VasilievTK,\VasilievAV} have received
increased attention in recent years due to their appearance within the  AdS/CFT  correspondence \refs{\HaggiManiRU,\KonsteinBI,\SundborgWP,\MikhailovBP,\SezginRT,
\KlebanovJA,\GiombiWH,\GiombiVG,\HenneauxXG,
\CampoleoniZQ,\GaberdielAR,\GaberdielPZ,\KochCY,\DouglasRC,\CastroCE,\GaberdielWB}.
These theories are more complex than  the usual supergravity approximation  in AdS/CFT where  only a finite number of fields are kept. However they are more tractable than  the  full string theory in AdS with infinite towers of massive string excitations.

One interesting conjecture \KlebanovJA\ relates higher spin gravity in AdS$_4$ to
the $O(N)$ vector model in three dimensions.  Another \GaberdielPZ\ relates a higher
spin theory in AdS$_3$ to a certain two-dimensional conformal field theory
with enhanced $\Wc$-algebra symmetry.   Higher spin gravity also appears
in the context of a constructive derivation of holography for free field theory \refs{\KochCY,\DouglasRC}. 

Three-dimensional higher spin theories \BlencoweGJ\ are considerably simpler to work with than their higher dimensional cousins, for two reasons.  First, it is consistent to truncate the tower of higher spin fields to a finite set, namely to only include those with spin $s \leq N$.  Second, the action can be expressed very   conveniently as a Chern-Simons theory.  By contrast, in
higher dimensions the theory necessarily includes an infinite tower of higher spin fields, and an action principle has only been written down very recently \refs{\BoulangerDD,\DoroudXS}.

We  focus here on the simplest higher spin theory in three-dimensions, which consists of
a metric  coupled to a spin-3 field.   Ordinary gravity in AdS$_3$
has left and right moving Virasoro algebras as its  asymptotic symmetry
group \BrownNW.  BTZ black holes carry nonzero  Virasoro zero-mode charges,
corresponding to mass and angular momentum.  In the spin-3 theory the
Virasoro algebras are enhanced to $\Wc_3$ algebras \refs{\HenneauxXG,\CampoleoniZQ}. From  the $\Wc_3$ algebra we have additional spin-3 charges that commute with the mass and
angular momentum, and so it is natural to seek generalized black hole
solutions that carry these charges. The purpose of this work is
to find and study these solutions.\foot{A candidate higher spin black hole solution in four dimensions was found in \DidenkoTD.}

To do so, we first need to extend previous discussions of the spin-3
theory and to develop some elements of the AdS/CFT dictionary.  In particular, previous work has considered solutions with  certain asymptotically AdS$_3$ boundary conditions, but as we discuss, black holes carrying spin-3 charge
are expected to have different asymptotic behavior.  To allow for
charged black holes, we show how to extend the boundary conditions;
in the AdS/CFT context this corresponds to allowing for nonzero
sources coupled to the spin-3 boundary currents.  Besides allowing
for black holes, this step is necessary for developing a framework to compute correlation functions involving the spin-3 operators.

With  suitably generalized boundary conditions in hand, we proceed
to find charged black hole solutions, which can be written down in a simple and explicit manner.   Evaluating their thermodynamic properties involves some novel elements compared to the familiar case
of ordinary gravity.  The most convenient gauge for finding solutions
is not one in which there is a manifestly smooth horizon, and so we cannot directly impose the usual condition of regularity of the Euclidean signature solution to find the temperature.   We instead employ an approach based on evaluating the holonomies of the Chern-Simons gauge field, and
show that this is consistent with the first law of thermodynamics and with smoothness of the horizon in the linearized limit.
Similarly, we do not base our computation of the black hole entropy
on computing the area of the event horizon, but instead by demanding
consistency with the first law.   The output of this computation is
an explicit formula for the black hole entropy as a function of the
black hole's mass, angular momentum, and spin-3 charges.

The BTZ black hole can be thought of as contributing to a finite
temperature partition function, and the fact that the BTZ entropy takes
the form of Cardy's formula allows it to be matched to the asymptotic
number of states of a dual boundary CFT \StromingerEQ.  Similarly, our spin-3 black holes
contribute to a partition function that includes chemical potentials for
the spin-3 charge.  The black hole entropy then yields a prediction for
the asymptotic growth of states at fixed charge in any candidate dual CFT
with $\Wc_3$ symmetry.   It will of course be very interesting to see
whether this prediction can be verified.

\newsec{ Spin-3 gravity and SL(3,R) Chern-Simons theory}

We begin by briefly reviewing relevant aspects of spin-3 gravity in three dimensions, along with the embedding of the BTZ black hole in this theory.

\subsec{Chern-Simons action}

It was discovered   long  ago that Einstein gravity with a negative cosmological constant can be reformulated as a SL(2,R)$\times $ SL(2,R) Chern-Simons theory \refs{\WittenHC,\AchucarroVZ}.  It was shown in  \refs{\BlencoweGJ,\BergshoeffNS} that a SL(N,R)$\times $ SL(N,R) Chern Simons theory corresponds to Einstein gravity coupled to $N-2$ symmetric tensor fields of spin $s=3,4, \ldots, N$.

In the following we will only consider the $N=3$ case, i.e., SL(3,R)$\times $ SL(3,R)
Chern-Simons theory, which corresponds to  spin-3 gravity in three dimensions with a negative cosmological constant.
Our conventions follow the recent paper \CampoleoniZQ, apart from a few differences that we
shall note explicitly.

 The action is
\eqn\abb{ S = S_{CS}[A] - S_{CS}[\Ab]}
where
\eqn\ac{  S_{CS}[A] ={k\over 4\pi} \int\! \tr \left(A\wedge dA +{2\over 3} A\wedge A \wedge A\right)}
The 1-forms $A$ and $\Ab$ take values in the Lie algebra of SL(3,R). An explicit representation of the eight generators $L_i,~  i=-1,0,+1$ and $W_j,~ j=-2,-1,\cdots,+2$, as well as our conventions, is given in
appendix A.  The Chern-Simons level $k$ is related to the
Newton constant $G$ and AdS$_3$ radius $l$ as
\eqn\ad{k ={l\over 4G}}
We henceforth set $l=1$.

The Chern-Simons equations of motion correspond
to vanishing field strengths,
\eqn\ae{ F = dA + A\wedge A =0~,\quad \Fb= d\Ab+\Ab\wedge \Ab =0 }
To relate these to the spin-3 Einstein equations we introduce a vielbein
$e$ and spin connection $\omega$ as
\eqn\af{ A = \omega+e ~,\quad \Ab= \omega -e }
Expanding $e$ and $\omega$ in a basis of 1-forms $dx^\mu$,
the spacetime metric $g_{\mu\nu}$  and spin-3 field $\varphi_{\mu\nu\gamma}$
are identified as
\eqn\ag{ g_{\mu\nu} = {1\over 2} \tr (e_\mu e_{\nu} )~,\quad \varphi _{\mu\nu\gamma} ={1\over 9 \sqrt{-\sigma}} \tr ( e_{(\mu} e_\nu e_{\gamma)})}
where $\varphi_{\mu\nu\gamma}$ is totally symmetric as indicated. Here
$\sigma$ is a parameter that can be absorbed into the normalization
of the $W_n$ generators; we assume $\sigma <0$ but otherwise keep it arbitrary.
For vanishing spin-3 field, $\varphi=0$, the flatness conditions \ae\
can be seen to be equivalent to Einstein's equations for the metric
$g_{\mu\nu}$ with a torsion free spin-connection.  More generally,
we find equations describing a consistent coupling of the metric to
the spin-3 field.

Acting on the metric and spin-3 field, the SL(3,R) $\times$ SL(3,R)  gauge  symmetries of the Chern-Simons theory turn into diffeomorphisms  along with
spin-3 gauge transformations (the Chern-Simons gauge transformation also include frame rotations, which leave the metric and spin-3 field invariant).    Under diffeomorphisms, the metric
and spin-3 field transform according to the usual tensor transformation rules.   The spin-3 gauge transformations are less familiar, as
they in general act nontrivially on both the metric and spin-3 field.
It is worth noting, though, that if we ignore the spin-3 gauge invariance,
then we can view the theory as a particular diffeomorphism invariant theory
of a metric and a rank-3 symmetric tensor field.

\subsec{Asymptotically AdS$_3$ boundary conditions}

To motivate the form of asymptotically AdS$_3$ boundary conditions, it is
helpful to note that the metric of global AdS$_3$, written in Fefferman-Graham coordinates,
\eqn\ah{ds_{AdS}^2 = d\rho^2 - \big(e^\rho +{1\over 4} e^{-\rho}\big)^2 dt^2 +\big( e^\rho -{1\over 4} e^{-\rho}\big)^2 d\phi^2 }
is obtained from the connections
\eqn\ai{\eqalign{ A_{AdS} & = \big( e^\rho L_1 +{1\over 4} e^{-\rho}L_{-1}\big) dx^+ + L_0 d\rho \cr
\Ab_{AdS}  & =   -\big( e^\rho L_{-1} +{1\over 4} e^{-\rho}L_{1}\big) dx^- - L_0 d\rho}}
where  $x^\pm = t \pm \phi$, and $\phi \cong \phi+2\pi$.    Introducing
\eqn\aj{ b(\rho) = e^{\rho L_0} }
we can write
\eqn\ak{\eqalign{ A_{AdS} & = b^{-1}\big(  L_1 +{1\over 4} L_{-1}\big)b dx^+ + b^{-1} \p_\rho b d\rho  \cr
\Ab_{AdS}  & =   -b \big(  L_{-1} +{1\over 4} L_{1}\big)b^{-1} dx^-  +b \p_\rho b^{-1} d\rho  }}
In \CampoleoniZQ, an asymptotically AdS$_3$ connection $A$ is taken
to obey $A_-=0$,  $A_\rho = b^{-1}(\rho) \p_\rho b(\rho)$, and
\eqn\al{   A- A_{AdS} \sim {\cal O}(1)~~~ {\rm as }~~ \rho \rt \infty}
The analogous condition for $\Ab$ is also imposed.

Upon using the freedom to make gauge transformations (see \CampoleoniZQ\ for a proof), asymptotically
AdS$_3$ connections can be taken to have the form
\eqn\am{\eqalign{ A & = b^{-1} a(x^+) b + b^{-1} db~,\quad
\Ab  = b \ab(x^-) b^{-1} + b db^{-1} }}
with $b$ given by \aj, and\foot{Our  $\Lc$ and $\Lcb$  are defined with opposite sign relative to those in \CampoleoniZQ.  With the sign convention used here, black holes will have $\Lc, \Lcb \geq 0$.}
\eqn\an{\eqalign{ a(x^+)&=\Big( L_1 -{2\pi \over k} \Lc(x^+) L_{-1} +{\pi \over 2k\sigma} \Wc(x^+) W_{-2}\Big)dx^+ \cr
\ab(x^-)&= -\Big( L_{-1} -{2\pi \over k} \Lcb(x^-) L_{1} +{\pi \over 2k\sigma} \Wcb(x^+) W_{2}\Big)dx^- }}

Given these connections, the asymptotic symmetry algebra is obtained
by finding the most general gauge transformation that preserves
the asymptotic conditions.    The functions appearing in the connections,  $\Lc(x^+)$ etc., transform under these gauge transformations.  Expanding in modes, one thereby arrives at two
copies of the classical $\Wc_3$ algebra.  $\Lc$ and $\Wc$ are identified with the stress tensor and spin-3 current. The modes of $\Lc$ by themselves form a Virasoro algebra with the standard Brown-Henneaux central charge.     We again refer to \CampoleoniZQ\ (and \HenneauxXG) for the details.

\subsec{BTZ black hole}

Before turning to generalizations, it is useful to review a few aspects of  the standard BTZ black hole in this formulation; see \refs{\WittenHC,\WittenKT,\BanadosGG} for relevant background on the Chern-Simons description and the BTZ solution.
For a BTZ black hole of mass $M$ and angular momentum $J$ we take
\eqn\ao{ \Lc = {M-J \over 4\pi}~,\quad \Lcb = {M+J \over 4\pi} }
Setting $\Wc = \Wcb=0$, we have the connections
\eqn\ap{\eqalign{ A& =\big( e^\rho L_1 -{2\pi \over k} e^{-\rho}\Lc L_{-1}\big)dx^+ + L_0 d\rho \cr
 \Ab& =-\big( e^\rho L_{-1} -{2\pi \over k} \Lcb e^{-\rho}L_{1}\big)dx^- - L_0 d\rho}}
Using  \af\ and \ag\ yields the BTZ metric in the form
\eqn\aq{\eqalign{  ds^2& =  d\rho^2 +{2\pi  \over k} \Big( \Lc (dx^+)^2 +\Lcb (dx^-)^2 \Big)-\Big( e^{2\rho}+\left({2\pi \over k}\right)^2\Lc \Lcb e^{-2\rho}\Big) dx^+ dx^-}}
As above, writing $x^\pm = t\pm \phi$, the angular coordinate $\phi$ is taken to have $2\pi$ periodicity.
The black hole entropy is
\eqn\ar{ S = {A_H\over 4G } = 2\pi  \Big( \sqrt{2\pi k\Lc}+ \sqrt{2\pi k \Lcb}~\Big)        }
Expressed in terms of the Brown-Henneaux central charge,
\eqn\as{ c = {3 \over 2G } = 6k }
we have the well-known result that the entropy takes the form of
Cardy's formula.

Going to Euclidean signature, $x^+ \rt z$ and $x^- \rt -\zb$,
demanding the absence of a conical singularity at the horizon imposes the periodicity conditions
\eqn\at{ (z,\zb) \cong (z+2\pi\tau, \zb + 2\pi \taub)}
where the modular parameter is given by
\eqn\au{ \tau  = {ik \over 2}{1\over \sqrt{2\pi k \Lc}}~,\quad
\taub= {-ik\over 2}{1\over \sqrt{2\pi k \Lcb} }}
The Hawking temperature $T=1/\beta$, and the angular velocity of the horizon $\Omega$ are then given by
\eqn\av{  \tau = {i\beta +i\beta \Omega \over 2\pi}~,\quad \taub = {-i\beta +i\beta \Omega \over 2\pi}}
Note that $J$ and $\Omega$ should be continued to pure imaginary
values in order to obtain a real Euclidean section.

In the framework of the AdS$_3$/CFT$_2$ correspondence, the BTZ black hole is to be thought of as contributing to the partition function
\eqn\av{ Z(\tau,\taub)  = \Tr_{AdS}~e^{4\pi^2 i \tau \hat{\Lc} - 4\pi^2 i \taub \hat{\Lcb}} = \Tr_{CFT} (q^{L_0-{c\over 24}} \bar q^{\bar L_0 -{c\over 24}})}
where $\hat{\Lc}$ and $\hat{\Lcb}$ now denote the Virasoro zero mode operators:  $2\pi \hat{\Lc} = L_0$, $2\pi \hat{\Lcb} = \overline{L}_0$.
In particular, the contribution to the  partition function  is given
by $e^{-I}$, where $I$ is the Euclidean Einstein-Hilbert action supplemented by appropriate boundary terms.  The classical black hole
geometry  gives a good approximation to the full partition function
at high temperatures and for $k\gg 1$,
 i.e. for a CFT with very  large central charge.

\newsec{Generalized boundary conditions}

A CFT with $\Wc_3$ symmetry has, by definition,  a dimension $(3,0)$
primary field $\Wc$.  Similarly, assuming $\Wc_3$ symmetry on the anti-holomorphic side we
have a dimension $(0,3)$ primary $\Wcb$.    We can add to the CFT
Lagrangian source terms for these operators,
\eqn\ba{ I ~\rt ~ I - \int\! d^2x \Big( \mu(x)\Wc(x) + \mub(x) \Wcb(x)\Big)}
Since the operators have scaling dimension $3$, they are non-renormalizable in two-dimensions, which makes it unclear at this stage whether we can make sense of the path integral for finite values of $\mu$ and $\mub$. On the other hand, we should certainly be able to obtain sensible results at the level of perturbation theory in the sources.

We now want to give a prescription for computing in the bulk in the
presence of these source terms.  According to the rules of the AdS/CFT
correspondence, we should associate the sources with boundary conditions for the spin-3 field in the bulk.
In the last section we reviewed
the boundary conditions for asymptotically AdS$_3$ solutions, but it's clear that these now need to be generalized.  Since the spin-3 operators
are  irrelevant in the renormalization group sense, adding them to the action changes the UV structure of the would-be CFT; the bulk
analog of this statement is that the geometry will no longer asymptote to the same AdS$_3$ geometry as in the undeformed theory.

Without further ado, we propose to consider connections of the form
\eqn\ama{\eqalign{ A & = b^{-1} a(x^+) b + b^{-1} db~,\quad
\Ab  = b \ab(x^-) b^{-1} + b db^{-1} }}
where, as above, $b=e^{\rho L_0}$, and now
\eqn\bb{\eqalign{ a&= \Big(L_1 -{2\pi \over k}\Lc L_{-1}+ {\pi \over 2k\sigma}\Wc W_{-2}  \Big) dx^+ \cr
& \quad +\Big( \mu W_2 + w_1 W_1 + w_0 W_0 + w_{-1}W_{-1} + w_{-2}W_{-2}+  \ell L_{-1} \Big)dx^-  \cr
 \ab&= -\Big(L_{-1} -{2\pi \over k}\Lcb L_{1}+ {\pi \over 2k\sigma}\Wcb W_{2}  \Big) dx^- \cr
&\quad -\Big( \mub W_{-2}  + \wb_{-1}W_{-1} + \wb_0 W_0 + \wb_1 W_1+ \wb_{2}W_{2}+  \ellb L_{1} \Big)dx^+  }}
As shown in \CampoleoniZQ,  by performing a gauge transformation one can always arrange
for $a_+$ and $\ab_-$ to take the above form.  The formulas for $a_-$ and $\ab_+$ are dictated by the structure of the field equations, as seen in
the next section.
At this stage, $a$ contains $8$ unspecified functions, $(\Lc, \Wc, \mu, w_1, w_0, w_{-1}, w_{-2}, \ell)$, all of which are allowed to depend on both $x^+$ and $x^-$, but not $\rho$; the analogous statement holds for $\ab$. We claim that the functions $\mu$ and $\mub$ should be identified with the sources in \ba.  To verify this claim we will compare the bulk field equations evaluated on this ansatz to the Ward identities in the CFT in the presence of spin-3 sources.

\newsec{Ward identities}

In this section we translate the bulk field equations for spin-3
gravity into Ward identities for the stress tensor and spin-3 current. We will focus on the $A$-connection; the treatment of $\Ab$
proceeds along exactly the same lines.  This computation will establish
the AdS/CFT dictionary in the presence of sources for the spin-3 operators. See \BanadosNR\ for an analogous computation in ordinary gravity.

\subsec{Bulk field equations}

According to \ama, the connection $A$ is gauge equivalent to $a$, and therefore the field equations can be written as
\eqn\ca{ da + a\wedge  a =0}
Plugging the ansatz \bb\ into \ca\ and solving iteratively, we find
\eqn\cb{\eqalign{ w_1 & = -\p_+ \mu \cr
w_0 &= {1\over 2}\p_+^2 \mu  -{4\pi \over k}\Lc \mu  \cr
w_{-1}& = -{1\over 6} \p_+^3 \mu +{4\pi \over 3k }  \p_+ \Lc \mu  +{10 \pi \over 3k} \Lc \p_+ \mu  \cr
w_{-2} & ={1\over 24}\p_+^4 \mu - {4\pi \over 3k }  \Lc \p_+^2 \mu - {7 \pi \over 6 k }\p_+ \Lc  \p_+ \mu   -{ \pi \over 3k } \p_+^2 \Lc \mu +{4\pi^2 \over k^2} \Lc^2 \mu  \cr
\ell  & = {4\pi \over k}  \Wc \mu}}
and that $\Lc$ and $\Wc$ are subject to
\eqn\cc{ \eqalign{ \p_- \Lc &= -3\Wc  \p_+ \mu  -2   \p_+ \Wc \mu \cr
\p_- \Wc & = {\sigma k \over 12 \pi} \p_+^5 \mu -{2\sigma \over 3}  \p_+^3 \Lc \mu-3\sigma   \p_+^2 \Lc \p_+ \mu -5\sigma   \p_+ \Lc \p_+^2 \mu -{10 \sigma \over 3}  \Lc  \p_+^3 \mu\cr
 &\quad +{64\pi \sigma \over 3k} \Lc  \p_+ \Lc  \mu +{64\pi \sigma \over 3k }  \Lc^2 \p_+ \mu }}
Any solution of these equations is a solution of \ca.  Note that
$\mu$ can be chosen freely.

\subsec{Stress tensor Ward identity}

 Our conventions and the operator product expansion for the stress tensor  and the spin-3 current are reviewed in appendix B. In particular, the  OPE for the stress tensor and the spin-3 primary is
\eqn\cd{ T(z) \Wc(0) \sim {3\over z^2} \Wc(0) +{1\over z}\p \Wc(0)+\cdots }
Due to the singular terms in the OPE, adding the source term
-$\int\! d^2z~ \mu(z,\zb) \Wc$ to the CFT action causes the stress tensor to pick up a $\zb$ dependence.    To characterize this we compute
\eqn\ce{ \p_{\zb} \langle T(z,\zb) \rangle_\mu }
where $\langle \cdots \rangle_\mu$ denotes an insertion of $e^{\int \mu \Wc}$ inside the expectation value.   Expanding in powers of $\mu$, and using
\eqn\cf{ \p_{\zb} \left({1\over z}\right)  = 2\pi \delta^{(2)}(z,\zb)}
we find
\eqn\cg{ {1\over  2\pi}  \p_{\zb} \langle T(z,\zb)\rangle_\mu  = -\langle 3  \Wc(z,\zb)\p_z \mu(z,\zb) +2  \p_z \Wc(z,\zb)\mu(z,\zb)  \rangle_\mu }
Converting to Lorentzian signature, this agrees with \cc\ under the expected identification $\Lc = -{1\over 2\pi }T$. It is quite convenient that the stress tensor corresponds precisely to a single term in the connection, in contrast to what one has in  the metric formulation where there are also contributions from boundary counterterms \BalasubramanianRE.

\subsec{Spin-3 Ward identity}

The OPE between two spin-3 currents is\foot{More precisely, this is to be regarded as the ``classical" large $k$ version.  At finite $k$ the
coefficients of the terms quadratic in $\Lc$ are changed, and the $\Lc^2$ operator needs a normal ordering prescription \ZamolodchikovWN; see Appendix B.  Expanding in $1/k$, such corrections would correspond to quantum effects in the
bulk.  We also note that symmetry algebras containing nonlinear terms have appeared before in the AdS/CFT context \refs{\HenneauxIB,\LapanJX,\KrausVU}.}
\eqn\ch{\eqalign{ \Wc(z) \Wc(0) &\sim  {5 \sigma k \over \pi^2} {1\over z^6}-{10\sigma \over \pi} {1\over z^4}\Lc  -{5\sigma \over \pi}{1\over z^3}\p \Lc -{3\sigma \over 2\pi}{1\over z^2} \p^2\Lc +{\sigma \over 3\pi} {1\over z} \p^3 \Lc \cr
&\quad +{32 \sigma \over 3k}{1\over z}\Lc  \p \Lc +{32 \sigma \over 3k}{1\over z^2} \Lc^2  }}
Our sign convention  differs from that in \ZamolodchikovWN, but is more convenient for our purposes.
To compare this to formulas in \CampoleoniZQ, recall that for a symmetry transformation  generated by $\chi(z)\Wc(z)$,  Noether's
theorem yields the  variation $\delta_\chi O$ as
\eqn\ci{\delta_\chi \Oc =2\pi  {\rm Res}_{z\rt 0} \left[\chi(z) \Wc(z) \Oc(0)\right] }
Applying this to $\Oc = \Wc$, we have
\eqn\cj{\delta_\chi \Wc = -{\sigma \over 3} \Big( 2\chi \Lc''' +9 \chi' \Lc''+15 \chi'' \Lc'+10 \chi'''\Lc -{k\over 4\pi} \chi^{(5)} -{64\pi \over k} (\chi \Lc \Lc' +\chi' \Lc^2) \Big)}
which is the same as (4.20b) of \CampoleoniZQ, taking into account
the sign flip in the relative definitions of $\Lc$.

Proceeding in the same manner as led to \cg, we now find
\eqn\ck{\eqalign{ -\p_{\zb} \Wc(z) &= {\sigma k \over 12\pi} \p_z^5 \mu -{10 \sigma \over  3} \p_z^3 \mu \Lc - 5 \sigma \p_z^2 \mu \p_z \Lc -3 \sigma \p_z \mu \p_z^2 \Lc -{2\over 3}\sigma  \mu \p_z^3\Lc\cr
&\quad +{64 \pi \sigma \over 3k} \mu \Lc \p_z \Lc +{64 \pi \sigma \over 3k} \p_z \mu \Lc^2   }}
This agrees with the second line in \cc\ after converting from Euclidean to Lorentzian signature.

\subsec{Comments}

Having satisfied the Ward identities, we now know that we have a consistent holographic dictionary for computing correlation functions of the stress tensor and spin-3 current.  In particular, we have established that in \bb\ we should interpret $\Lc$ and $\Wc$ as
the stress tensor and spin-3 current, and $\mu$ as the source for the spin-3 current.  A source for the stress tensor can be introduced
by allowing the coefficient of $L_1$ in the expansion of $a$ in \bb\
to be a nontrivial function of $x^\pm$.

The solution \cb-\cc\ can be used as the starting point for computing AdS/CFT correlation functions of the stress tensor and spin-3 current. The equations \ce\ can be solved iteratively as a power series in $\mu$, and the functional coefficient of the $\mu^n$ term gives the
correlation function of $\Lc/\Wc$ with $n$ additional insertions of the spin-3 current.  Of course, in solving \cc\ one needs to invert
$\p_-$, which requires imposing boundary conditions, which in the
AdS/CFT correspondence are usually imposed at the Poincare horizon
of the bulk geometry.

Since the Ward identities are equivalent to the OPEs, which can in turn be used to derive the symmetry algebra obeyed by the mode operators, we can view the above computations as a derivation of the
$\Wc_3$ algebra obeyed by the bulk higher spin theory.  This is to be compared to the derivation in \CampoleoniZQ, which in our language
set $\mu=0$ and then considered the algebra of gauge transformations
preserving specified asymptotic boundary conditions.   Consistency
demands that these two approaches yield the same result, as we have
indeed verified.

Before proceeding, we should emphasize that the connections \ama-\bb,
at finite $\mu$ and $\mub$, correspond to solutions that do not asymptote
to an AdS$_3$ geometry of radius $l=1$. The metric is easily seen to contain terms growing like $e^{4\rho}$, while asymptotic  $l=1$  AdS$_3$ metrics grow as $e^{2\rho}$.   Instead, the metric asymptotes to AdS$_3$ with a
different radius, namely $l=1/2$.    As noted above, this change in the asymptotic structure is to be expected from the AdS/CFT dictionary given that we are
adding to the CFT action an irrelevant dimension $3$ operator.  To
further address whether this is sensible, we now turn to the physical
interpretation of such  solutions, and will see that a consistent
picture emerges.

\newsec{Black holes with spin-3 charge}

\subsec{Solutions}

To obtain black hole solutions we consider the ansatz \ama-\bb,
but with all functions independent of $x^\pm$.  From \cb\ (and the analogous formulas for $\ab$) we see that solutions take the form
\eqn\da{\eqalign{ a&= \Big(L_1 -{2\pi \over k}\Lc L_{-1} +{\pi \over 2k\sigma} \Wc  W_{-2} \Big)dx^+  \cr
& \quad +\mu \Big(W_2 -{4\pi \Lc \over k}W_0 +{4\pi^2 \Lc^2 \over k^2} W_{-2} +{4\pi \Wc \over k}L_{-1}\Big)dx^- \cr
 \ab&= -\Big(L_{-1} -{2\pi \over k}\Lcb L_{1} +{\pi \over 2k\sigma} \Wcb  W_{2} \Big)dx^- \cr
 & \quad -\mub \Big(  W_{-2} -{4\pi \Lcb \over k}W_0 +{4\pi^2 \Lcb^2 \over k^2} W_{2} +{4\pi \Wcb \over k}L_{1}\Big) dx^+ }}
The corresponding connections $(A,\Ab)$ are
\eqn\db{\eqalign{ A&= \Big(e^\rho L_1 -{2\pi \over k}\Lc  e^{-\rho}L_{-1} +{\pi \over 2k\sigma} \Wc e^{-2\rho} W_{-2} \Big)dx^+  \cr
& \quad +\mu \Big(e^{2\rho} W_2 -{4\pi \Lc \over k}W_0 +{4\pi^2 \Lc^2 \over k^2} e^{-2\rho} W_{-2} +{4\pi \Wc \over k}e^{-\rho} L_{-1}\Big)dx^-  +L_0 d\rho  \cr
 \Ab&= -\Big(e^\rho L_{-1} -{2\pi \over k}\Lcb e^{-\rho} L_{1} +{\pi \over 2k\sigma} \Wcb e^{-2\rho} W_{2} \Big)dx^- \cr
 & \quad -\mub \Big( e^{2\rho}  W_{-2} -{4\pi \Lcb \over k}W_0 +{4\pi^2 \Lcb^2 \over k^2} e^{-2\rho} W_{2} +{4\pi \Wcb \over k}e^{-\rho} L_{1}\Big) dx^+ -L_0 d\rho  }}
The metric and spin-3 field are extracted using \ag; we will just display the metric:
\eqn\dc{\eqalign{  ds^2& = d\rho^2
-4\sigma\Big( \mu e^{2\rho} dx^- +{\pi \over 2k \sigma}\Wcb e^{-2\rho} dx^-+{4\pi^2 \over k^2}\mub \Lcb^2 e^{-2\rho} dx^+\Big)\cr
 & \quad\quad\quad \quad \times \Big( \mub e^{2\rho} dx^+ +{\pi \over 2k \sigma}\Wc e^{-2\rho} dx^++{4\pi^2 \over k^2}\mu \Lc^2 e^{-2\rho} dx^-\Big) \cr
&- \Big( e^\rho dx^+ -{2\pi \over k}\Lcb e^{-\rho}dx^- +{4\pi \over k} \mub \Wcb e^{-\rho} dx^+\Big)\Big( e^\rho dx^- -{2\pi \over k}\Lc e^{-\rho}dx^+ +{4\pi \over k} \mu \Wc e^{-\rho} dx^-\Big) \cr
&-{\sigma \over 3}\left({4\pi \over k}\right)^2 \big( \mu \Lc dx^- + \mub \Lcb dx^+\big)^2}}
A few immediate comments are in order.  Since we are taking $\sigma<0$, we see that the metric has the correct signature.  For
$\mu=\mub=\Wc=\Wcb=0$, the metric is that of a rotating BTZ black hole, but it becomes deformed once the spin-3 field is turned on.
We also see very explicitly that for $\mu \mub \neq 0$ the metric grows like
$e^{4\rho}$, and so asymptotes to AdS$_3$ with radius $l=1/2$,  as discussed above.

\subsec{Interpretation and specification of parameters}

Just as the Euclidean BTZ solution can be thought of as contributing to the partition function \av, we now wish to think of the higher spin black hole solutions as contributing to a generalized partition function which includes chemical potentials conjugate to the spin-3 currents. An analogous situation occurs when the bulk theory consists of gravity along with pure Chern-Simons gauge fields; the AdS/CFT correspondence has been well developed for that case \refs{\DijkgraafFQ,\KrausNB}.     With this in mind we consider the partition function
\eqn\dd{ Z(\tau,\alpha,\taub,\alphab)= \Tr~ e^{4\pi^2i [ \tau \hat{\Lc}+ \alpha \hat{\Wc } -\taub \hat{\Lcb}-\alphab\hat{\Wcb}]} =\Tr_{CFT} q^{L_0-{c\over 24}} u^{W_0} \bar q^{\bar L_0-{c\over 24}} \bar u^{W_0}}
On the CFT side the partition function can be expanded in terms of generalized characters of the $\Wc_3$ algebra. As discussed briefly in Appendix B, not much is known about these characters.

As will be seen, the potentials $(\alpha,\alphab)$ are related to $(\mu,\mub)$ as
\eqn\de{ \alpha = \taub \mu~,\quad \alphab= \tau \mub }
In the BTZ solution, the stress tensor $\Lc$ is related to the
modular parameter $\tau$ according to \au, so that the solution
is labelled by either $\tau$ or $\Lc$ (and similarly for the
barred quantities).    By the same token, in our solution \db, after
imposing the periodicity conditions \at\ we need to relate
$(\Lc, \Wc)$ to $(\tau, \alpha)$ in order to obtain solutions with the
same free parameters as appear in the partition function \dd.
From the CFT standpoint these assignments are understood as expectation values, according to
\eqn\df{\eqalign{ \Lc &=\langle \hat{\Lc}\rangle= -{i\over 4 \pi^2} {\p \ln Z \over \p \tau}~,\quad \Wc =\langle \hat{\Wc}\rangle=  -{i\over 4 \pi^2}{\p \ln Z \over \p \alpha} }}
The problem for us is to specify the physical conditions that determine $(\Lc, \Wc)$ in terms of  $(\tau, \alpha)$.

For a spin-3 black hole solution, we would like to impose the following 3 conditions:

\medskip

\noindent
{\bf 1.} The euclidean geometry is smooth and the spin-3 field is nonsingular at the horizon.  In the BTZ solution,
after fixing $\tau$ the value of $\Lc$ is obtained by demanding that
the Euclidean geometry close off smoothly at the location of the
Lorentzian event horizon.   This is of course well known to
yield the correct relation between the black hole mass and Hawking temperature.

\medskip

\noindent
{\bf 2.} In the limit $\mu\to 0$ the solution goes smoothly over to the BTZ black hole. In particular, we want that  $\Wc\to 0$ and the black hole entropy becomes \ar.

\medskip

\noindent
{\bf 3.} The following constraint should be
satisfied by any consistent assignment of $(\Lc,\Wc)$.  From \df, we see that if $(\Lc,\Wc)$ are to arise from an underlying partition function then they should obey the integrability condition
\eqn\dg{ {\p \Lc \over \p \alpha} = {\p \Wc \over \p \tau}}
Another way to say this is that \dg\ needs to be satisfied in order
that the thermodynamic quantities assigned to the black hole will obey the first law of thermodynamics.

\medskip

As we shall now discuss, it is not obvious that all three conditions can be satisfied at the same time.

\subsec{Nonrotating solution}

The basic complications can be seen in the nonrotating case, where we take $\tau = -\taub = i\beta$, $\mub=-\mu$, $\Lcb=\Lc$, and  $\Wcb =-\Wc$.  We also set $\sigma =-1$ for the moment.  The metric  \dc\  becomes,

\eqn\dkg{\eqalign{ds^2 & = d\rho^2  - \Big\{ \big( 2\mu e^{2\rho}+{\pi \over k}\Wc e^{-2\rho}-{8\pi^2 \over k^2}\mu \Lc^2 e^{-2\rho}\big)^2  +\big(e^\rho-{2\pi \over k}\Lc e^{-\rho}+{4\pi \over k}\mu \Wc e^{-\rho}\big)^2 \Big\} dt^2 \cr
&  +\Big\{ \big(e^\rho+{2\pi \over k}\Lc e^{-\rho}+{4\pi \over k}\mu \Wc e^{-\rho}\big)^2  +4\big( \mu e^{2\rho}+{\pi \over 2k}\Wc e^{-2\rho}+{4\pi^2 \over k^2}\mu \Lc^2 e^{-2\rho}\big)^2  \cr
&\quad +{4\over 3} \left(4\pi \over k\right)^2 \mu^2 \Lc^2\Big\} d\phi^2  }}

 The horizon occurs where $g_{tt}$ vanishes, and
since the radial dependence of the metric is simply $d\rho^2$,
we need $g_{tt}$ to have a double zero at the horizon in order for
the Euclidean time circle to smoothly pinch off. One solution is given by $\Wc=0$ and $\mu=0$, which is  the nonrotating  BTZ black hole. There is a second solution provided
\eqn\conwla{k+ 32 \mu^2 \pi (\mu \Wc- \Lc) =0}
is satisfied.
The temperature is then determined by demanding the absence of a conical singularity at the horizon, which is located at $e^{\rho_+}= \sqrt{(2\pi\Lc -4\pi \mu \Wc )/k}$,
 \eqn\dkj{\beta = 2\pi \left.\sqrt{2 \over -g_{tt}''}~\right|_{\rho=\rho_+}= {2\sqrt{2} \pi k \mu\over \sqrt{(3 k - 32\pi  \Lc \mu^2 ) ( 16 \pi \Lc \mu^2 -k)}}}
For the nonrotating solution \de\ becomes $\alpha=-i\beta\mu$. The two equations \conwla\ and \dkj\ can be solved  in terms of $\Lc$ and $\Wc$ to give

\eqn\dko{\eqalign{
\Lc & = {k \over 64  \pi }{\beta^2 \over \alpha^2} \left(\sqrt{1-64\sigma \pi^2 \alpha^2 /\beta^4}-5\right)\cr
\Wc & = -{k \over 64 \pi }{\beta^3 \over \alpha^3}  \left(\sqrt{1-64\sigma \pi^2 \alpha^2 /\beta^4}-3 \right) }}

It is easy to see from \dko\ that $\Lc$ and $\Wc$ diverge in the limit $\mu\to 0$. Furthermore the integrability condition \dg\ (and hence the first law of thermodynamics) is violated for \dko. Hence, while the metric  \dkg\ satisfies condition 1 of section 5.2, it violates conditions 2 and 3.

In the next section we will propose a different approach based on gauge invariant information of the Chern-Simons formulation.

\subsec{Holonomy condition}

Since Chern-Simons theory is a theory of flat connections, the gauge
invariant information is contained in the holonomies.   Let us therefore step back and consider the holonomies associated to the
BTZ black hole.   In general, the holonomy associated with
the identification $(z,\zb)\cong (z+2\pi \tau, \zb+2\pi \taub)$ is
\eqn\dh{\eqalign{ H &=b^{-1} e^{\omega}b ~,\quad \omega =  2\pi ( \tau a_+ - \taub a_-)\cr
\Hb &=b e^{\omb}b^{-1}~,\quad \omb =  2\pi ( \tau \ab_+ - \taub \ab_-)  }}
If we evaluate these for the BTZ black hole, using
the relations \au, we find that   $\omega$ and $\omb$ have
eigenvalues $(0,2\pi i, -2\pi i)$. We take this as the gauge invariant information associated with the Euclidean time circle.  In particular, demanding that these eigenvalues are realized leads to the correct
relation between $\Lc$ and $\tau$, namely \au.

We now come to the key point:  for the generalized  spin-3  black hole we will continue to demand that $\omega$ and $\omb$ have eigenvalues
$(0,2\pi i, -2\pi i)$, the motivation being that this represents
the gauge invariant characterization of a smooth horizon.
As further --- compelling in our view --- evidence in favor of this proposal, we now show that it leads to a solution of the integrability condition \dg,
and coincides with the smoothness condition for a linearized spin-3 field on the BTZ background.

\subsec{Integrability condition}

Rather than working directly with  eigenvalues, it is convenient to impose the equivalent conditions
\eqn\di{\eqalign{  \det(\omega) &=0~,\quad \Tr(\omega^2)+8\pi^2 =0 \cr  \det(\omb)&=0~,\quad \Tr(\omb^2)+8\pi^2 = 0}}
Using \da, these conditions are explicitly
\eqn\dj{\eqalign{ 0&= 2048 \pi^2 \sigma^2 \alpha^3\Lc^3+576 \pi \sigma k \tau^2 \alpha \Lc^2 +864 \pi \sigma k \alpha^2 \tau \Wc \Lc +864 \pi \sigma k \alpha^3 \Wc^2-27 k^2 \tau^3 \Wc \cr
0 & =256 \pi^2 \sigma \alpha^2 \Lc^2  -24 \pi k \tau^2 \Lc-72 \pi k \tau  \alpha \Wc -3k^2\cr }}
together with the same formulas with unbarred quantities replaced by their barred versions. We have replaced $\mu$ by $\alpha$ using \de.

To verify the integrability conditions we proceed as follows.
First solve the second equation for $\Wc$ and differentiate with
respect to $\tau$ to get an expression for ${\p \Wc \over \p\tau}$
in terms of  ${\p \Lc \over \p\tau}$.
Next, insert the solution for $\Wc$ into the first equation, and
then differentiate with respect to $\alpha$, and solve to get an
expression for ${\p \Lc \over \p \alpha}$.   Similarly, differentiate
with respect to $\tau$ to get an expression for  ${\p \Lc \over \p \tau}$.  Substituting the latter into our previous expression for ${\p \Wc \over \p\tau}$, we find that it precisely equals ${\p \Lc \over \p \alpha}$, which is the desired integrability condition.

\subsec{Linearized spin-3 solution}

We now look for a solution representing a smooth linearized spin-3 perturbation of
a non-rotating BTZ black hole.   We'll find that the smoothness condition is equivalent to imposing the condition \di\ on the holonomy.

The connections corresponding to non-rotating BTZ are
\eqn\dsa{\eqalign{ A_{BTZ}&=  \left(e^{\rho} L_1  -{2\pi \over k}\Lc e^{-\rho} L_{-1}  \right) dx^+ +L_0 d\rho  \cr
\Ab_{BTZ}&=  -\left(e^{\rho} L_{-1}  -{2\pi \over k}\Lc e^{-\rho} L_{1}  \right) dx^- -L_0 d\rho
}}
From
our general result \db, a linearized solution with $\mub=-\mu$ and
$\Wcb=-\Wc$ is obtained by adding
\eqn\dsb{\eqalign{ \delta A & = {\pi \over 2k\sigma}\Wc e^{-2\rho} W_{-2} dx^+ +\mu \left(e^{2\rho}W_2 -{4\pi \over k}\Lc W_0 +{4\pi^2 \over k^2}\Lc^2 e^{-2\rho}W_{-2}\right)dx^- \cr
\delta \Ab & = {\pi \over 2k\sigma}\Wc e^{-2\rho} W_{2} dx^- +\mu \left(e^{2\rho}W_{-2} -{4\pi \over k}\Lc W_0 +{4\pi^2 \over k^2}\Lc^2 e^{-2\rho}W_{2}\right)dx^+
 }}
where we're dropping terms of order $\mu \Wc$.
To linear order the metric is
\eqn\dsc{ ds^2 = d\rho^2 -\left(e^\rho -{2\pi \over k} \Lc e^{-\rho}\right)^2 dt^2 +\left(e^\rho +{2\pi \over k} \Lc e^{-\rho}\right)^2 d\phi^2 }
with a horizon at
\eqn\dsd{ e^{\rho_+} = \sqrt{2\pi \Lc \over k}}
The corrections to the metric are quadratic in $(\mu,\Wc)$, and
so are zero in the linearized approximation.

The holonomy  around the Euclidean time direction is,
\eqn\dscaa{\omega = 2\pi(\tau A_+ -\taub A_-)  }
with $\tau$ given
by its usual BTZ expression,
\eqn\dscaab{\tau = {ik\over 2}{1\over \sqrt{2\pi k \Lc} } }
The eigenvalues of $\omega$, to first order in $(\mu,\Wc)$, are computed to be
\eqn\dscab{ i\delta,~~ \pm 2\pi i -{i\delta \over 2} }
with
\eqn\dscac{
\delta = \sqrt{2\pi \over -\sigma k \Lc^3}\left({ 64 \pi \sigma \Lc^2 \mu +3k\Wc \over 12}\right) }
Demanding that these eigenvalues are unchanged from their BTZ values
therefore fixes $\Wc$ in terms of $\mu$ as
\eqn\dscad{ \Wc= -{64 \pi \sigma \Lc^2 \over 3k}\mu}
We now want to see whether this condition coincides with demanding smoothness of the perturbation.

The spin-3 field, normalized here as
\eqn\dsca{ \varphi= \varphi_{\alpha\beta\gamma}dx^\alpha dx^\beta dx^\gamma = \Tr (e_\alpha e_\beta e_\gamma)dx^\alpha dx^\beta dx^\gamma }
is found to have nonzero components: $\varphi_{\phi\phi\phi}$, $\varphi_{\phi \rho\rho}$, $\varphi_{\phi tt}$.
Of particular interest is the behavior near the horizon.
We compute
\eqn\dsf{\eqalign{\varphi &= \varphi_{\alpha\beta\gamma} dx^\alpha dx^\beta dx^\gamma  \cr & =\left( -{24 \pi \Wc\over \sqrt{-\sigma} k}(\rho- \rho_+) + {4\pi \over \sqrt{-\sigma}k^2}(224 \pi \sigma \Lc^2 \mu +9k \Wc)(\rho-\rho_+)^2 + \cdots \right) d\phi dt dt    \cr
& \quad +\left( {16 \pi \sqrt{-\sigma} \Lc \over k} \mu + \cdots\right)d\phi d\rho d\rho \cr
&\quad +\Big( {4\pi \over \sqrt{-\sigma}k^2}(64 \pi \sigma \Lc^2 \mu -3k \Wc) + \cdots\Big)d\phi d\phi d\phi   }}
In general, this is singular at $\rho=\rho_+$.  Just as we require
$g_{tt}$ to have a double zero at the horizon, so too must $\varphi_{\phi tt}$.

In order to get a smooth solution we need to perform a gauge transformation.
Under a gauge transformation by parameter $\lambda$ we have
\eqn\dsg{ \delta_\lambda  A = d\lambda +[A,\lambda]}
and similarly for barred quantities.  Now, since the background
solution only involves $L$-type generators, and the trace of
any mixed bilinear combination vanishes, $\Tr L W =0$, it's clear
that the background metric, $g_{\mu\nu}={1\over 2} \Tr (e_\mu e_\nu)$, will be invariant under a gauge transformation with $\lambda$ made
up of purely $W$-type generators. These will just act on the spin-3
field.  So we consider
\eqn\dsh{\eqalign{\lambda &= \lambda_2 W_2 +\lambda_1 W_1 +\lambda_0 W_0 +\lambda_{-1} W_{-1} +\lambda_{-2} W_{-2} \cr
\lamb &= \lamb_2 W_2 +\lamb_1 W_1 +\lamb_0 W_0 +\lamb_{-1} W_{-1} +\lamb_{-2} W_{-2}}}
We want to preserve translation invariance along $(t,\phi)$, so
we take the parameters to depend only on $\rho$.   After a bit of
experimentation, the simplest option is
\eqn\dsha{\lambda_1 = -\lambda_{-1}=-\lamb_1 =\lamb_{-1} = f(\rho)}
with the rest of the parameters vanishing.
Acting on the background solution, this  induces nonzero components for $\varphi_{\phi\phi\phi}$, $\varphi_{\phi \rho\rho}$, $\varphi_{\phi tt}$, which are the same components as appear
 in the original linearized solution. Expanding around the horizon
we find
\eqn\dshb{\eqalign{ \delta_\lambda \varphi & = -192 \sqrt{-\sigma} \sqrt{2\pi^3 \Lc^3\over k^3} f(\rho_+) (\rho-\rho_+)d\phi dt^2 -192\sqrt{-\sigma} \sqrt{2\pi^3 \Lc^3\over k^3} f'(\rho_+)(\rho-\rho_+)^2 d\phi dt^2 \cr
& +24 \sqrt{-\sigma} \sqrt{2\pi \Lc \over k} f'(\rho_+) d\phi d\rho^2 }}
To cancel off the previous $(\rho -\rho_+)$ term in $\varphi_{\phi tt}$ we thus take
\eqn\dshc{ f(\rho_+) =\sqrt{k \over 2\pi \Lc^3}  {\Wc \over 8\sigma}  }

To achieve a smooth solution we still have to impose one more condition.
Near the horizon, the metric in the $(it,\rho)$ plane approaches the origin of flat space in polar coordinates provided the imaginary time direction is
identified with the correct periodicity.  This periodicity is determined from the result
\eqn\dsj{ {g_{tt}'' \over g_{\rho\rho} }\Big|_{\rho_+} = -{16 \pi \Lc \over k}}
Since the $\phi$ direction is effectively inert in this computation,
smoothness of the spin-3 field demands that we have the same ratio:
\eqn\dsk{  {\varphi_{\phi tt}'' \over \varphi_{\phi\rho\rho} }\Big|_{\rho_+} =- {16 \pi \Lc \over k}}
From \dshb\ we observe that the values computed from $\delta_\lambda \varphi$
obey this by themselves.  This means that the gauge transformation \dsha\ acting on the BTZ background
leads to a smooth solution, which is reassuring.  It also means that the linearized solution \dsf\  needs to satisfy this condition by
itself, so that the total solution, corresponding to $A=A_{BTZ}+\delta A + \delta_\lambda A$, will as well.  This condition is evaluated  to be
\eqn\dsl{ -{112\pi \Lc \over k} -{9 \Wc \over 2\sigma \Lc \mu} = - {16 \pi \Lc \over k}}
which yields
\eqn\dsm{\Wc = -{64 \pi \sigma \Lc^2 \over 3k}\mu}
Comparing with \dscad, we see precise agreement.   Thus, at the linearized
level smoothness is equivalent to the holonomy condition \di.

\subsec{Summary}

We have shown that the holonomy condition \di\ leads to a solution that satisfies conditions 2 and 3 of section 5.2, as well as condition 1 in the
linearized limit.   We take this as strong evidence that this is the physically correct condition to impose in order to define the spin-3 black holes.

This proposal does however have a surprising implication that we would
ultimately like to understand better in the future.
In the special case of the nonrotating solution with metric  \dkg, it is not hard to see that with  $\Lc$ and $\Wc$ determined by \dj, the time component of the metric $g_{tt}$ never vanishes.  This solution, therefore, does not
at first glance appear to be a black hole at all, but rather a wormhole with a second asymptotic region extending out to $\rho \rt -\infty$.
One may therefore question whether we have really satisfied condition 1.

We believe that the resolution of this issue has to do with the spin-3
gauge transformations, which we recall act nontrivially on the metric as well as the spin-3 field.  We already saw in our linearized analysis that
it was necessary to perform such a gauge transformation in order to exhibit
a manifestly smooth horizon.  More generally, we expect that the non-appearance of an event horizon in the fully nonlinear solutions is a consequence of our choice of gauge, and that an appropriate gauge transformation will restore it.   Here we should emphasize that one's usual intuition about the definition of an event horizon has to be modified in this context, due to the noninvariance of the metric under spin-3 gauge transformations.  This is a fascinating issue that deserves to be better understood, but for now we will proceed under the assumption that the
holonomy condition \di\ defines the spin-3 black hole.

\newsec{Black hole thermodynamics}

Having satisfied the integrability condition, we know that if we now
compute the entropy from the partition function it is guaranteed to be consistent with
the first law of thermodynamics.   For black holes in Einstein-Hilbert
gravity, we can of course directly compute the entropy in terms of the
area of the event horizon.  But in the present context we do not know
{\it a priori} whether the entropy is related to the area in this way, in
particular due to the nontrivial spin-3 field.   We instead base our entropy
computation on demanding adherence to the first law.  Since this approach may not be so familiar, in Appendix C we run through the logic in the familiar setting of the Reissner-Nordstrom black hole.

Expressed as a function of $(\Lc,\Wc)$ the entropy $S$  obeys the following thermodynamic relations:
\eqn\dk{ \tau= {i\over 4\pi^2}{\p S \over \p \Lc}~,\quad   \alpha={i\over 4\pi^2}{\p S \over \p\Wc} }
The analogous barred relations hold as well.  Since the entropy breaks
up into a sum of an unbarred piece plus a barred piece with identical structure, in the following we just focus on the unbarred part and add the
two parts at the end.

We can  use dimensional analysis to write the entropy in terms of
an unknown function of the dimensionless ratio $\Wc^2 /\Lc^3$.
With some foresight, it proves convenient to write
\eqn\dl{ S = 2\pi \sqrt{2\pi k \Lc} f(y) }
with
\eqn\dn{ y =  {27 k\Wc^2 \over - 64 \sigma \pi \Lc^3} }
Demanding agreement with the BTZ entropy imposes $f(0)=1$.  Using \dk\ and plugging  into the second line of \dj\  we arrive at the following differential equation
\eqn\dm{ 36y \big(2 - y\big)(f')^2+f^2-1=0}
This equation also implies  the first equation in \dj.
The solution with the correct boundary condition is
\eqn\do{ f(y) = \cos \theta~,\quad \theta = {1\over 6} \arctan \left({\sqrt{y(2-y)}\over 1-y}\right)}
 The physical range of $y$ is given by $0 \leq y \leq 2$, and we choose a branch of the arctangent such that $0 \leq \theta \leq {\pi \over 6} $.

Our final result for the entropy, including both sectors, is thus
\eqn\dpz{ S = 2\pi \sqrt{2\pi k \Lc} f\Big( {27 k\Wc^2 \over - 64 \sigma \pi \Lc^3}\Big)+ 2\pi \sqrt{2\pi k \Lcb} f\Big( {27 k\Wcb^2 \over - 64 \sigma \pi \Lcb^3}\Big) }
For small argument, $f(y)$ has the expansion
 \eqn\dq{ f(y) \sim 1 - {1\over 36}y -{35 \over 7776}y^2 -{1001 \over 839808}y^3 +\cdots }
 Near $y=2$ we have
 \eqn\dr{ f(y) \sim \sqrt{3\over 4} +{\sqrt{2}\over 12}\sqrt{2-y}+\cdots }
For given $\Lc$ we thus have a maximal spin-3 charge $\Wc$ given by
\eqn\ds{ \Wc_{max}^2 = {-128 \sigma \pi \over 27 k }\Lc^3}
As the maximal value is approached $\tau$ diverges according to \dk, corresponding to vanishing chiral temperature.  On the other hand, the entropy is finite, attaining a relative value of $\sqrt{3\over 4}$ compared to the entropy at $\Wc=0$.  This behavior is to be contrasted with that of the BTZ black hole, or its charged generalizations with respect to bulk
U(1) gauge fields \KrausWN.  In those cases, whenever the temperature of one chiral
sector goes to zero, so too does the entropy associated with that sector.
For extremal BTZ black holes the entropy is carried entirely by the sector
at nonzero temperature, whereas here a zero temperature sector can contribute to the entropy.

Given the entropy, $\tau$ and $\alpha$ can be computed using \dk, and from there
we compute the partition function according to
\eqn\dt{\ln Z =S+ 4 \pi^2 i \left( \tau \Lc +\alpha\Wc -\taub \Lcb -\alphab \Wcb \right)}
This partition function should match the asymptotic behavior of
the partition function of any candidate CFT dual to the higher spin theory in the bulk.

\newsec{Discussion}

We conclude with a few assorted observations and open questions.

The natural extension of our work is to consider charged black
holes in the general spin-$N$ gravity theory.   The large $N$ limit
is of particular interest given that the explicit duality proposal in \GaberdielPZ\ involves  large $N$.  Similarly, the bulk field content
arising in \GaberdielPZ\ involves massive scalar fields, and it would
be interesting to see how they interact with the black holes.  In particular, they could serve as a useful physical probe of the putative event horizon.

A key step in our logic involved using the holonomies to fix the black hole parameters in a manner consistent with the first law of thermodynamics, and smoothness in the linearized limit.    This was shown to work by direct computation, but it would be very helpful to have a more conceptual understanding of this procedure.  Another important issue to be resolved concerns the existence of a smooth event horizon in the nonlinear regime.  We conjectured that there exists
a spin-3 gauge transformation that can be used to exhibit a manifestly nonsingular horizon, and we hope that this can be demonstrated explicitly.\foot{This conjecture has now been established in \AmmonNK.}   More generally, greater understanding of the role of spin-3 gauge transformations will be very helpful for gaining a better physical understanding of the theory.

Some subtleties associated with black holes in higher spin gravity were discussed in \CastroCE.   For sufficiently large $N$ (at fixed
$k$), the number of linearized states in the bulk appears to exceed the Cardy bound, and hence also exceeds the number of states associated with black holes.   These issues do not arise in the present work, as we are considering the regime $k \gg N$ where the number of linearized bulk states is parametrically smaller.

The generalized partition functions of CFTs with $\Wc_3$ symmetry
have apparently not been studied in the literature.  Our black hole
entropy result suggests that a universal formula for the asymptotic behavior exists, at least at large $k$, and  it would be very interesting to see whether this can be established directly in CFT.

\vskip .3in

\noindent
{ \bf Acknowledgments}

\vskip .3cm

 This work was supported in part by NSF grant PHY-07-57702. We are grateful to Eric Perlmutter for discussions, and  to Gerard Watts for a useful correspondence.

\appendix{A}{SL(3,R) generators}

As in  \CampoleoniZQ, we use the following basis of SL$(3,R)$ generators
\eqn\za{ \eqalign{ L_1 & = \left(\matrix{0&0&0 \cr 1&0&0 \cr 0&1&0}\right),\quad L_0= \left(\matrix{1&0&0 \cr 0&0&0 \cr 0&0&-1}\right),\quad  L_{-1} = \left(\matrix{0&-2&0 \cr 0&0&-2 \cr 0&0&0}\right)\cr & \cr
W_2 &= 2\sqrt{-\sigma} \left(\matrix{0&0&0 \cr 0&0&0 \cr 1&0&0}\right),\quad W_1 = \sqrt{-\sigma} \left(\matrix{0&0&0 \cr 1&0&0 \cr 0&-1&0}\right),\quad  W_0 = {2\over 3}\sqrt{-\sigma} \left(\matrix{1&0&0 \cr 0&-2&0 \cr 0&0&1}\right)\cr & \cr
W_{-1} &= \sqrt{-\sigma} \left(\matrix{0&-2&0 \cr 0&0&2 \cr 0&0&0}\right),\quad W_{-2} = 2\sqrt{-\sigma} \left(\matrix{0&0&4 \cr 0&0&0 \cr 0&0&0}\right) }}
The parameter $\sigma$ will always be taken to be negative, but is otherwise left unspecified; it can be set to any desired value by rescaling the $W$ generators.
The generators obey the following commutation relations
\eqn\zb{\eqalign{ [L_i ,L_j] &= (i-j)L_{i+j} \cr
[L_i, W_m] &= (2i-m)W_{i+m} \cr
[W_m,W_n] &= {\sigma \over 3}(m-n)(2m^2+2n^2-mn-8)L_{m+n} }}
and trace relations
\eqn\zc{ \eqalign{ \tr ( L_0 L_0) &= 2~,\quad   \tr ( L_1 L_{-1} ) = -4 \cr   \tr ( W_0 W_0 )  &= -{8\sigma \over 3}~,\quad \tr ( W_1 W_{-1} )  = 4\sigma ~,\quad \tr ( W_2 W_{-2} )  = -16\sigma   }}
All other traces involving a product two generators vanish.

\appendix{B}{$\Wc_3$ algebra}
 The $\Wc_3$ algebra \ZamolodchikovWN\ has in addition to the stress energy tensor $T(z),$ a primary $W(z)$ of conformal weight 3.  These operators have mode expansions on the plane
\eqn\apba{
T(z)= \sum_n L_n z^{-n-2}, \quad W(z) =\sum_n W_n z^{-n-3}}
Their operator product expansions are given by
\eqn\apbb{\eqalign{T(z)T(w) &= {c/2 \over (z-w)^4 }+ {2 \,T(w)\over (z-w)^2} + {\partial_w T(w) \over z-w }+\cdots \cr
T(z)W(w) &=  {3\, W(w) \over (z-w)^2}+ {\partial_w W \over z-w} +\cdots \cr
W(z)W(w)&={c/3\over (z-w)^6} + {2\,T(w)\over (z-w)^4}+ {\partial_w T(w) \over (z-w)^3} + {1\over (z-w)^2} \left( 2 \beta \Lambda(w) + {3\over 10}\, \partial_w^2 T(w) \right)\cr
& \quad + {1\over z-w} \left( \beta \,\partial \Lambda (w) + {1\over 15}\,\partial_w^2 T(w) \right) }}
where the parameter $\beta$ is given by
\eqn\apbc{\beta={16\over 22+5c}}
and the operator $\Lambda$ can be defined as follows
\eqn\apbd{\Lambda= : T(w) T(w): - {3\over 10}\, \partial_w^2 T(w)}
A ``classical" version of this algebra is obtained by taking the large $c$ limit and disregarding the normal ordering.  It is this version of the algebra that is reproduced in the dual gravitational theory. In particular, the $\Wc \Wc$ OPE agrees with \ch\  upon setting $\sigma = -1/10$, $c=6k$, and identifying $T=-2\pi \Lc$, $W = 2\pi \Wc$.

One can define a highest weight representation based on a highest weight  state $ \mid\phi_{h,w}\rangle$ which satisfies
\eqn\highestw{\eqalign{&L_0 \mid \phi_{h,w}\rangle = h \mid \phi_{h,w}\rangle  \cr
&W_0 \mid \phi_{h,w}\rangle = w \mid \phi_{h,w}\rangle  \cr
&L_n \mid \phi_{h,w}\rangle =0, \quad  W_n \mid \phi_{h,w}\rangle =0, \quad {\rm for} \quad n>0}}
One considers the Verma module $V(h,w,c)$ associated with the highest weight vector $\phi_{h,w}$ as spanned by the linearly independent basis vectors
\eqn\vermaa{\eqalign{&L_{-m_1} L_{-m_2} \cdots L_{-m_n} W_{-n_1} W_{-n_2} \cdots W_{-n_k} \mid \phi_{h,w}\rangle\cr
& {\rm with} \;\;\; m_1\geq m_2\geq \cdots \geq m_n > 0, \quad  n_1\geq n_2\geq \cdots \geq n_k > 0 }}
We can consider the character of the Verma module $V$. For generic values of $c$ where there exist  no null submodules in the Verma module, the character has the following form.
\eqn\vermab{\chi_V(q) = \tr_V( q^{L_0-{c\over 24} }) = {q^{h -{c\over 24}} \over \Big(\prod_{n>0} (1- q^n)\Big)^2}}
Since the zero modes $W_0$ and $L_0$ commute, it is natural to consider  a generalized form of the character
\eqn\vermac{\chi_V(q,u) =  \tr_V( q^{L_0-{c\over 24} } u^{W_0})   }
However not much is known about these characters. In principle, $W_0$ can be expressed in Jordan normal form as an upper triangular matrix and the contribution  to \vermac\ for generic values of $c$ can be calculated \BouwknegtNG. However even at level $2$ this involves solving a quintic and can only be done numerically. The result does not seem to be illuminating.\foot{We thank Gerard Watts for a useful correspondence regarding these matters.} As far as we know,  the modular properties and asymptotic growth formulas for the number of states have not been calculated.

\appendix{C}{Black hole thermodynamics from integrability}

In determining the thermodynamics properties of our spin-3 black hole we employed a method based on an integrability condition and the first law of thermodynamics.   To explain the logic of this procedure it is useful to
illustrate it with a familiar example and show how we recover standard results, such as the area law for the entropy.

Our example is the Reissner-Nordstrom solution in D=3+1,
\eqn\xxa{\eqalign{ ds^2 & = -{(r-r_+)(r-r_-) \over r^2}dt^2 +  {r^2 \over (r-r_+)(r-r_-)}dr^2 + r^2 d\Omega^2 \cr
A_\mu dx^\mu & = -\left({Q\over r} - {Q\over r_+}\right) dt    }}
with (setting $G_N=1$)
\eqn\xxb{ r_\pm =  M \pm \sqrt{M^2 - Q^2}}
The area law gives the entropy: $S=A_H/4 = \pi r_+^2$.

The first step is to demand a smooth Euclidean solution.  Demanding the
absence of a conical singularity at $r=r_+$ requires the identification
$t \cong t + i\beta$, with the inverse Hawking temperature $\beta$ given by
\eqn\xxc{ \beta = {4\pi \over -g_{tt}'(r_+)} = {4\pi r_+^2 \over r_+ -r_-}}
Since the Euclidean time circle shrinks to zero at $r_+$, smoothness of the
gauge field requires that its holonomy vanish there.  This has already been enforced in \xxa\ by choosing the appropriate additive constant in $A_t$.
These smoothness conditions are the analog of our holonomy conditions for the spin-3 black hole.

An important quantity is the asymptotic value of $A_t$:
\eqn\xxd{ \mu  = A_t\big|_{r=\infty} = { Q\over r_+}}
It is well known that for the Reissner-Nordstrom solution $\mu$ plays the role of the chemical potential conjugate to the charge $Q$, but to maintain the parallel with our spin-3 problem we want to avoid making this assumption just yet; this interpretation will instead emerge from the analysis.

Now we come to the main issue, which is determining the black hole thermodynamic quantities
without assuming the area law for the entropy, or using the Gibbons-Hawking
relation to the Euclidean action, neither of which are immediately applicable in the spin-3 case.  It is conceptually helpful --- though not necessary --- to think of the black hole free energy as being obtained from a statistical mechanical partition function of the form
\eqn\xxe{ Z(\beta,\alpha) = {\rm Tr} ~e^{ - \beta H - \alpha Q } }
Given the partition function, the free energy is defined as $ F = -{1\over \beta} \ln Z$.   In the thermodynamic limit, which is all that we are considering here,  we think of \xxe\ as being dominated by states of a particular energy and charge, and so we have
\eqn\xxf{Z(\beta,\alpha) = e^{S} e^{ - \beta E - \alpha Q } }
and then the entropy $S$ can be extracted as
\eqn\xxg{  S = \ln Z + \beta E +\alpha Q}
The  energy and charge are determined by differentiating \xxe,
\eqn\xxh{ E(\beta,\alpha) = - {\p \ln Z \over \p \beta}~,\quad Q(\beta,\alpha) = - {\p \ln Z \over \p \alpha} }
The relations \xxh\ can be used to pass back and forth between $(E, Q)$ and $(\beta, \alpha)$.  For instance, the entropy $S$ can be expressed in terms of either set of parameters.

To apply this to the black hole case we need to specify how the parameters $(\beta, \alpha)$ appearing in the partition function are related to parameters appearing in the black hole solution.  It is of course very familiar that $\beta$ should be equated with the inverse Hawking temperature \xxc, and we will simply assume that here.   To determine $\alpha$  we will use the integrability condition implied by \xxh,
\eqn\xxi{ { \p E \over  \p\alpha } = {\p Q \over \p \beta} }
Since we know that $E$ and $Q$ are given by the black hole mass and charge, and $\beta$ by the inverse temperature, we can regard \xxi\ as an equation fixing $\alpha$,  and the solution is
\eqn\xxj{ \alpha =- {4\pi  r_+ Q \over r_+ -r_-} = -\beta \mu }
Given how $\alpha$ appears in  \xxe, we have thus derived that $\mu$ is indeed the standard chemical potential.  This solution for $\alpha$ is the analog of \de\ for the spin-3 case.

The partition function can now be found by integrating \xxh, which yields $\ln Z = -\pi r_+^2$.    Finally, the entropy is computed from
\xxg\ as $S=\pi r_+^2$, which reproduces the area law result.

These steps can be carried out for any charged black hole, and our spin-3 case is just a slightly more complicated example.  Once again, the virtue of this approach is that we are only basing our conclusions on consistency with the first law of thermodynamics, and not making any assumptions about a relation to the area law or the Euclidean action.

\listrefs
\end